\documentclass[cits]{PoS}

\usepackage[T1]{fontenc}
\usepackage{graphicx}
\usepackage{epsfig}
\usepackage{comment}
\usepackage{amssymb}
\usepackage[centertags]{amsmath}

\def\beq{\begin{equation}}
\def\eeq{\end{equation}}
\def\be{\begin{eqnarray}}
\def\ee{\end{eqnarray}}

\title{Charge asymmetry of top quarks}

\ShortTitle{Charge asymmetry of top quarks}

\author{\speaker{Paola Ferrario}\\%
        IFIC  (UVEG - Consejo Superior de Investigaciones Cient\'ificas)\\
        E-mail: \email{paola.ferrario@ific.uv.es}}

\author{Germ\'an Rodrigo\\
        IFIC  (UVEG - Consejo Superior de Investigaciones Cient\'ificas)\\
        E-mail: \email{german.rodrigo@ific.uv.es}}

\abstract{The LHC is a promising machine to discover new physics in the top sector. There are several models that predict the existence of heavy colored resonances decaying to top quarks in the TeV energy range. The production of such resonances might generate a sizable charge asymmetry of top versus antitop quarks. At the Tevatron, a $2\sigma$ discrepancy with the SM prediction for the forward--backward asymmetry has been found, boosting a renewed interest for this kind of models. We consider a toy model with general flavour dependent couplings of the resonance to quarks, of both vector and axial-vector kind and investigate the possible constraints on the coupling space from the measurement of the asymmetry and the invariant mass distribution at the Tevatron. Then, we define a central asymmetry in a specific kinematic region and investigate its signature at the LHC as well as its statistical significance, for exclusive processes.}

\FullConference{XVIII International Workshop on Deep-Inelastic Scattering and Related Subjects\\
		 April 19 -23, 2010\\
		 Convitto della Calza, Firenze, Italy}

\begin{document}

\section{Introduction}

The achievement of $7$ TeV in the centre-of-mass energy at the LHC has started off an exciting era for particle physics. An integrated luminosity of $1$ fb$^{-1}$ is planned for the first $18$ to $24$ months of running, where precise measurements of the Standard Model (SM) will be performed as well as hints of new physics possibly unveiled. The LHC is a top machine: of
% the cross section for the production of $t\bar t$ pairs is around $100$-$200$ pb, which means that
the order of $10^5$ top antitop quark pairs will be produced already in the first running, with $1$ fb$^{-1}$, which is a larger sample than at the Tevatron. With such a huge top statistic, it is natural investigating the reach of top production in order to find signals of new physics at the LHC. Several models predict the existence of heavy colored resonances decaying to top-antitop quark pairs that might be observed at the LHC, like axigluons in chiral color models, different kinds of massive gauge bosons in coloron and top color models or Kaluza Klein excitations in extra dimensional models. These resonances should be detectable in top-antitop quark events, particularly in those 
models where the coupling of the new gauge bosons to the third 
generation is enhanced with respect to the lighter fermions. Their existence modifies the $q\bar q\rightarrow t\bar t$ production cross-section, while gluon-gluon fusion to top quarks stays, at first order, 
unaltered, because a pair 
of gluons do not couple to a single extra resonance 
in this kind of models. 
%For example, 
% the asymmetric chiral color model \cite{Cuypers:1990hb}
% allows the existence of three axigluon vertices, which are forbidden 
% in the usual chiral color model by parity, 
% but exclude gluon-gluon-axigluon vertices as well. 
% Models in extra warped dimensions, where KK modes
% can be single produced, have been constructed \cite{Agashe:2007jb}, 
% but in the conventional and more extended extra dimensional 
% models, a single KK gauge field does not couple to two SM gauge bosons 
% at leading order by orthonormality of field profiles \cite{Dicus:2000hm}. The new interaction between quarks and the extra resonance reads, in the most general case:
% \be
% \mathcal{L}_{\rm res.}\,\equiv\, i g_s\bar\psi_q\gamma^\mu(g_V^q+g_A^q\gamma_5) R^a_\mu T_a\psi_q\, ,
% \ee
% where each quark with flavor $q$ can couple with different strength to the colored massive resonance $R^a$.

% Detecting such resonances with masses in the TeV region is not an easy task, because the top quarks are very energetic. This means that they are very boosted, thus their decay products are highly collimated. On one side, this makes the identification of the jets difficult, due to a high level of overlapping. On the other side, it is easier to misinterpret a $b$ quark as a light quark, since the two vertices that are the characteristic signal of the $b$ quark are very close to each other and they could be detected as a single jet \cite{Baur:2008uv}.

The natural signature of these resonances is a peak in the invariant mass distribution of the top-antitop quark pair located at the mass of the new 
resonance. % Colored resonances are fairly broad:
% $\Gamma_G/m_G = {\cal O}(\alpha_s) \sim 10\%$.
% Present lower bounds on their mass are about $1$~TeV.
% The latest exclusion limit by CDF \cite{masslimits} at 95\% C.L. is 
% $260$~GeV$ < m_G < 1.250$~TeV for axigluons and flavor-universal colorons
% (with $\cot \theta=1$ mixing of the two $SU(3)$). 
However, asymmetries can be an alternative way of revealing 
these resonances. Some of those exotic gauge bosons, such as the 
axigluons, might generate already at tree-level a charge asymmetry through 
the interference with the $q\bar q \to t \bar t$ 
SM amplitude \cite{Antunano:2007da, FBmodels}.
% The most stringent lower bounds on the mass of such new states
% are about $800$~GeV for the $W'$ and 
% $Z'$~\cite{Aaltonen:2008dn,Aaltonen:2009qu,:2007dia,d0ttbar}, 
% $1.2$~TeV for axigluons and flavor-universal colorons~\cite{Aaltonen:2008dn}, 
% and $600$~GeV for gravitons~\cite{graviton}. 
% Electroweak precision measurements rise the exclusion mass 
% region of the $Z'$ to above $3$~TeV in Randall-Sundrum 
% scenarios~\cite{Agashe:2003zs}.
% Those limits, however, should be taken with care as 
% they depend on the given model adopted to set that bounds, 
% although the numbers quoted above are quite similar across
% different analysis. 

\begin{figure}[htb]
  \begin{center}
    \includegraphics[width=14cm]{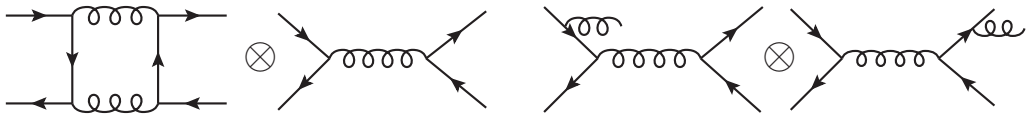}
  \end{center}\caption{\small{Graphs contributing to the QCD charge asymmetry in quark-antiquark production.}}
  \label{asym}
\end{figure}

QCD at tree level predicts that top-antitop quark pair production at hadron colliders is charge symmetric, namely the differential charge asymmetry, defined as:
\be A(\cos\theta)=\frac{N_t(\cos\theta)- N_{\bar t}(\cos\theta)}{N_t(\cos\theta)+ N_{\bar t}(\cos\theta)}\label{asym_c}\ee
vanishes for every value of $\theta$, where $N_{t(\bar t)}(\cos\theta)$ is the number of top (antitop) quarks produced at a certain angle $\theta$ with respect to the incoming quark. Nevertheless, an asymmetry is generated at $\mathcal{O}(\alpha_s^3)$ from the graphs shown in Fig.~\ref{asym} \cite{germantops}. 
% %
% \begin{figure}[htb]
%   \begin{center}
%     \includegraphics[width=7cm]{asymqqbar1b.ps} \includegraphics[width=7cm]{asymqqbar2b.ps}
%   \end{center}\caption{\small{Graphs contributing to the QCD charge asymmetry in quark-antiquark production.}}
%   \label{asym}
% \end{figure}
% %

%

The charge asymmetry coming from the real hard radiation has opposite sign compared to the soft and virtual corrections and the latter is always larger than the former. %The soft gluon radiation generates an infrared singularity that is canceled exactly by the one emerging in the double gluon exchange. 
On the other hand, $q g$ originated processes generate a contribution to the asymmetry much smaller than $q\bar q$, and $gg$ fusion is obviously symmetric.

% The way a charge asymmetry arises in QCD is analogous to what happens in Quantum Electrodynamics \cite{Berends:1973fd}. In a $e^+e^-\to \mu^+\mu^-$ annihilation, the cross section receives charge asymmetric contributions at $\mathcal{O}(\alpha^{3}_{\rm e. m.})$ from the same amplitudes as the ones shown in Fig.~\ref{asym} (where the gluon is replaced by a photon and the light and top quarks are replaced by electrons and muons respectively). The photon has charge conjugation $C=-1$. In those interferences, an odd number of photons appears and a minus sign under charge conjugation results, thus leading to an asymmetry in the $\mu^+$ and $\mu^-$ production. In QCD, the gluon has not a definite sign under charge conjugation and the above products of amplitudes contain both C-even and C-odd components. 
% 
% The graphics contributing to the asymmetry shown in Fig.~\ref{asym} ($1$) are related to the correspondent ones with the top exchanged with the antitop ($2$) by:
% \be  \sigma_{(1)}(t,\bar t)=-\sigma_{(2)}(\bar t,t)\,,\ee
% without taking into account the color factors, which are: \be C_1&\propto& d^2+f^2\nn
% C_2&\propto & d^2-f^2\,,\ee where $d^2\equiv d_{abc}d^{abc}$ and $f^2\equiv f_{abc}f^{abc}$. Thus, the asymmetry just selects the $d^2$ factor, meaning that only color-singlet quark-antiquark configurations contribute.  

\section{The charge asymmetry at the Tevatron}

At the Tevatron, the charge asymmetry is equivalent to the forward--backward asymmetry, from CP invariance. The forward--backward asymmetry of top quarks has already been measured
at the Tevatron proton-antiproton collider \cite{cdf,d0,newcdf}, at a center-of-mass energy of $\sqrt{s}=1.96$ TeV. The most recent value for the forward-backward asymmetry in the laboratory frame, extracted with a luminosity of $3.2$~fb$^{-1}$, is \cite{newcdf}: 
\beq
A_{\rm FB}^{p \bar p} = \frac{N_t (\cos \theta >0)-N_t (\cos \theta <0)}
{N_t(\cos \theta >0)+N_t(\cos \theta <0)}
= 0.193 \pm 0.065_{\, \rm stat.} 
\pm 0.024_{\, \rm syst.}\,,
\label{eq:cdf1}
\eeq
where $\theta$ is the angle between the top quark and the proton beam.
%and has to be compared with the one year old result:
%$A^{p\bar p} = 0.17 \pm 0.07_{\, \rm stat.} \pm 0.04_{\, \rm syst.}$, 
%with $1.9$~fb$^{-1}$~\cite{cdf}. The uncertainty of both measurements is still large, but 
%systematic errors have been improved considerably from 
%one measurement to another, 
%and statistical errors have decreased accordingly. 
%Moreover, it turns out to be quite interesting that the 
%uncertainty is still statistically dominated, 
%and hence significant improvements should be expected in the 
%near future. 

The total charge asymmetry generated at the Tevatron by QCD at NLO has been calculated to be \cite{Antunano:2007da,posgerm}:
\be A=\frac{N_t(y\geq 0)-N_{\bar t}(y\geq 0)}{N_t(y\geq 0)+N_{\bar t}(y\geq 0)}=0.051(6)\,,\label{asymNLO}\ee
where $y$ is the quark (antiquark) rapidity in the laboratory frame. Comparing the theoretical prediction 
\eqref{asymNLO} with the experimental result \eqref{eq:cdf1}, a discrepancy emerges of about two sigmas, that opens a window 
to the presence of new physics. 

In a general, model independent way, we consider heavy color-octet boson resonances decaying to $t\bar t$ with arbitrary vector and axial-vector couplings to quarks $g_V^q$ and $g_A^q$. In the Appendix of Ref.~\cite{Ferrario:2008wm} we list the differential cross section for $t\bar t$ production in the presence of such a resonance. A positive asymmetry 
can be generated if the term from the squared amplitude of 
the massive color-octet
%, which is proportional to 
%$8 g_V^q g_A^q g_V^t g_A^t c$, 
dominates over the term of 
the interference.
%, that is proportional to $2 g_A^q g_A^t c$. 
This is possible if the vector couplings are large 
enough. However, although the total 
cross section might still be compatible with the SM prediction
in that case, because the contribution of the excited gluon 
is suppressed by powers of its mass, the top-antitop quark invariant 
mass distribution might be enhanced considerably, 
%due to the factor 
% \be
% \left( (g_V^q)^2 + (g_A^q)^2 \right)\left( (g_V^t)^2 + (g_A^t)^2 \right)~,
% \ee
particularly for high values of the top-antitop quark invariant mass. In \cite{Aaltonen:2009iz}, the differential distribution in the invariant top-antitop quark mass is presented. The measurement is performed with $2.7$~fb$^{-1}$ of integrated luminosity. The invariant mass is divided in nine bins and the last one ($0.8-1.4~{\rm TeV}$)
% %
% \be
% \frac{d\sigma}{dM_{t\bar t}} (0.8-1.4~{\rm TeV}) &=&
% 0.068 \pm 0.032_{\, \rm stat.} \pm 0.015_{\, \rm syst.} \pm 0.004_{\, \rm lumi.}
% \label{mttbar}
% \ee
% %
is the most sensible to extra contributions beyond the SM at the TeV scale. The charge asymmetry and the invariant mass distribution probe different combinations of the vector and axial-vector couplings; therefore, by combining both limits, one can constrain complementary regions in the parameter space.

\begin{figure}[htb]
\begin{center}
\includegraphics[width=6cm]{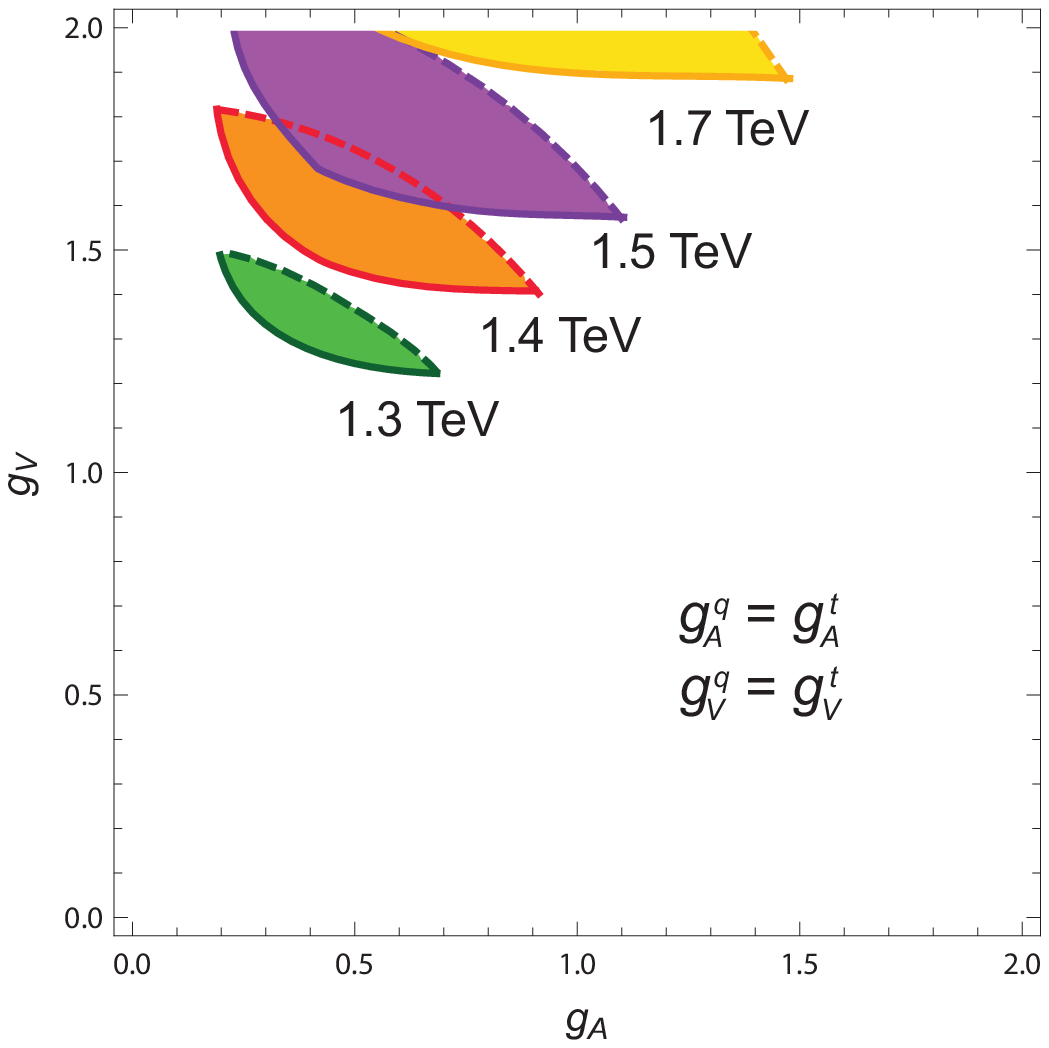}
\includegraphics[width=6.2cm]{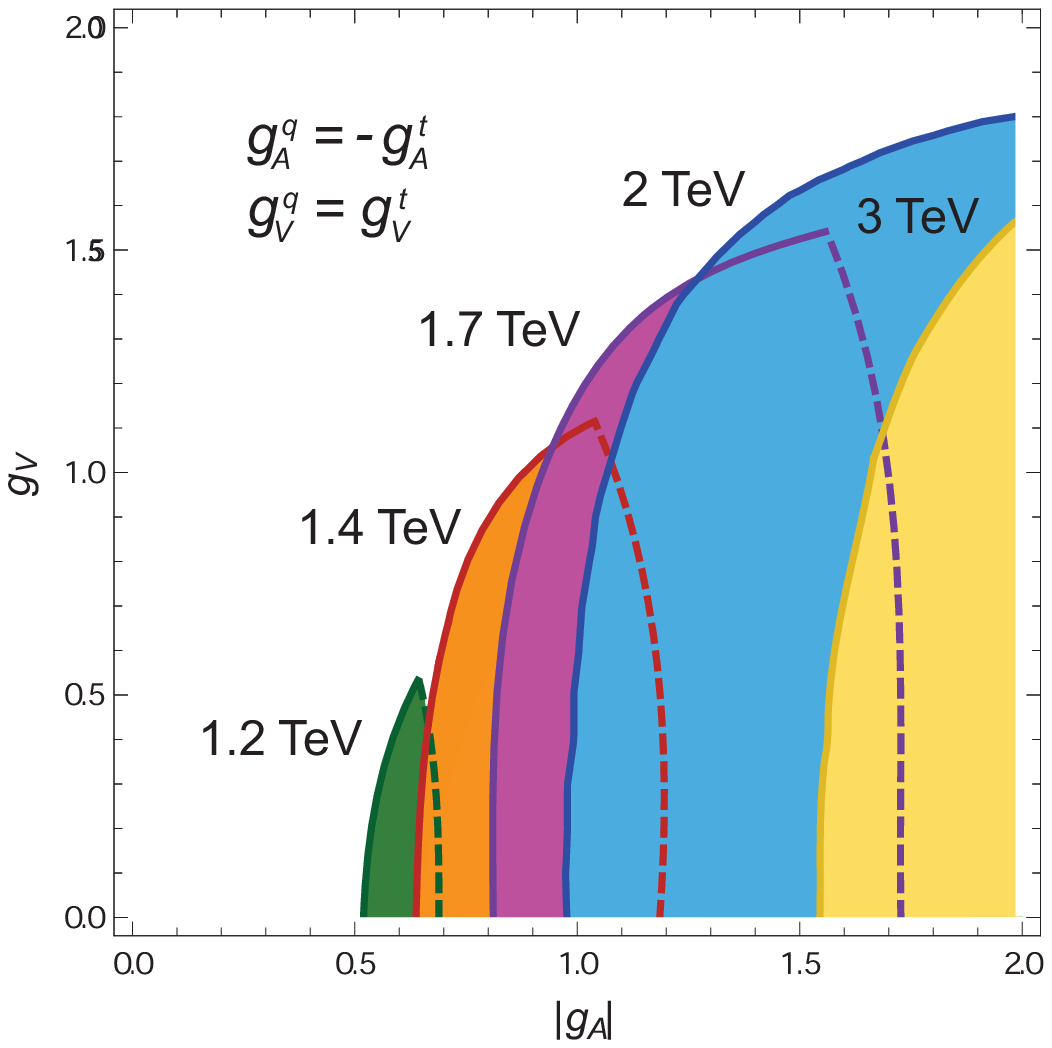}
\caption{Constraints on the vector and axial-vector couplings for different 
values of the resonance mass. In the flavour-universal case (left plot) the contours are shown at 95 \% C.L., while in the other case (right plot) at 90\% C.L.}
\label{constraints}
\end{center}
\end{figure}

Our results are shown in Fig. \ref{constraints} \cite{Ferrario:2009bz}, where, for 
a given value of the mass of the color-octet resonance, we provide 
the allowed region at $95$\% C.L. (for the flavor-universal scenario) and $90$\% (for the other one) in the $g_V-g_A$ plane.
%%%%%%%%%%%%%%%
%%%%%%%%%%%%%%%
The solid lines are obtained from the charge asymmetry, 
while the dashed lines are derived from the last bin of the 
invariant mass distribution. 
In the flavor-universal case, the allowed regions are quite constrained; indeed, at $90$\% C.L.
we do not find any overlapping region for any value of the 
color-octet mass, and future experimental measurements with 
higher statistics can shrink significantly, or even exclude 
completely the allowed regions at $95$\% C.L.. If the axial coupling has different sign for light and top quarks, $g_A^q=-g_A^t$, we find that, for $|g_A|<2$ and independently of the 
resonance mass, the region about 
\beq
(|g_A|-2.3)^2+|g_V|^2 \gtrsim 1.8^2
\eeq
is excluded at 90\% C.L. Furthermore, 
for fixed values of the vector and axial-vector couplings, 
the charge asymmetry sets a lower limit on the mass of the 
color-octet, while an upper bound can be set thanks to the 
invariant mass distribution.

\section{The charge asymmetry at the LHC}

The LHC is a proton-proton machine, thus its initial state is symmetric and the forward-backward asymmetry vanishes. However, it is still possible to find a charge asymmetry in selected kinematic regions. %as shown in Fig.~\ref{fig:boost}. 
%%%%%%%%%%%%%%%%%%%%%%%%%%%%%%%%%%%%%%%%%%%%%%%%%%%%%%%%%%%%%%%%%%%%%%
% \begin{figure}[htb]
% \begin{center}
% \includegraphics[width=8cm]{lhc1.eps} \\
% \includegraphics[width=8cm]{lhc2.eps}
% \caption{Boost from the center-of-mass quark--antiquark reference frame to the laboratory frame.  }
% \label{fig:boost}
% \end{center}
% \end{figure}
%%%%%%%%%%%%%%%%%%%%%%%%%%%%%%%%%%%%%%%%%%%%%%%%%%%%%%%%%%%%%%%%%%%%%%
%As said before, tops are produced more abundantly in the direction of the incoming quark. Momentum in protons is carried mainly by valence quarks (and not antiquarks). Thus, the boost from the centre of mass frame to the laboratory frame results in "squeezing" tops towards the beam axis, leaving more antitops in the central region, at low rapidities. 
We define a central asymmetry~\cite{Antunano:2007da}:
\be
A_C(y_C) = \frac{N_t(|y|\le y_C)-N_{\bar{t}}(|y|\le y_C)}
{N_t(|y|\le y_C)+N_{\bar{t}}(|y|\le y_C)}~.\label{eq:acyc}
\ee 
It obviously vanishes if the 
whole rapidity spectrum is integrated, while a non-vanishing 
asymmetry can be obtained over a finite interval of rapidity. %It is worth noticing that this asymmetry does not arise from any CP violating effect, but it is due to a restriction in the phase space. 
%According to what said above, a positive partonic asymmetry translates into a negative central asymmetry: an abundance of top quarks in the forward direction in the partonic cross section means that more antitop quarks are left in the central region in the laboratory frame.

At the LHC most of the events ($70\%$ at $7$ TeV) are gluon-gluon fusion, which are symmetric. However, it is possible to enhance the asymmetry through a cut on the top-antitop invariant mass, $m_{t\bar t} > m_{t\bar t}^{\rm{min}}$,
because that region of the phase space is more sensitive 
to the quark-antiquark induced events rather than the 
gluon-gluon ones.%, due to the behaviour of the parton distribution functions of the gluon and the light quarks. 

There are some models of resonances which do not produce a charge asymmetry at LO in the inclusive process. This happens if any of the axial-vector couplings vanishes, like, for instance, in some models of Kaluza-Klein gluon, where $g_A^q = 0$ for light flavours. Nonetheless, a charge asymmetry is generated in the exclusive $t\bar t$ + jet process: it is now a leading order effect and, besides the contribution of gluon bremsstrahlung from fermions, also the gluon emission from the exchanged particle contributes to the asymmetry.

\begin{figure}[!htb]
\begin{center}
\includegraphics[width=6cm]{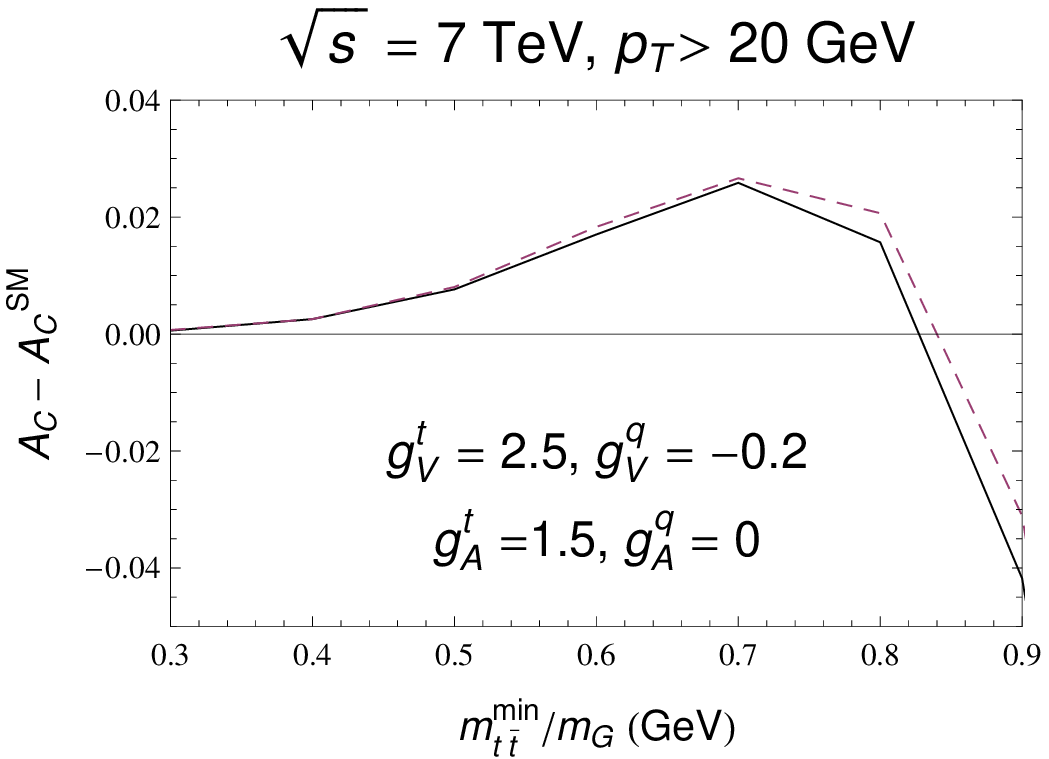}
\includegraphics[width=6cm]{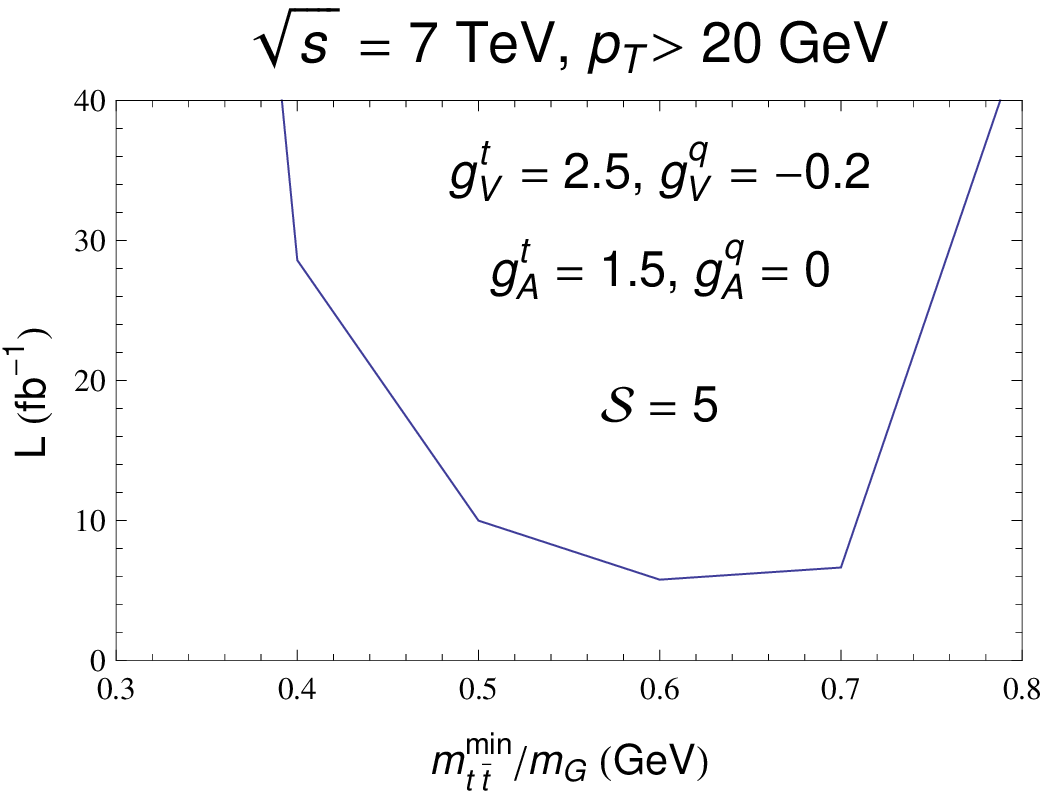}\\
\caption{Central charge asymmetry and luminosity to obtain a statistical 
significance $\mathcal{S}=5$ at the LHC, as a function of 
$m^{\mathrm{min}}_{t\bar t}$ for $\sqrt{s}=7$ TeV. 
The dashed line represent the contribution of the terms proportional to $d_{abc}^2$. 
$m_G=1.5$ TeV. \label{fig:7TeV}} 
\end{center}
\end{figure}

In Fig.~\ref{fig:7TeV} we show the result of our analysis for a scenario of a typical Kaluza-Klein gluon excitation \cite{Ferrario:2009ee}. We have calculated the central asymmetry and the luminosity needed to reach a statistical significance of $5$ as a function of the cut on the invariant top-antitop quark mass. We find that there is a minimum in this luminosity for relatively low cuts: this is a non-trivial result as it means that relatively low energetic top quarks can already generate a measurable asymmetry. This is an advantage, because very boosted top quarks are difficult to distinguish from jets initiated by light quarks. 

\section*{Aknowledgements}

This work is supported by Grant
No. FPA2007-60323, by Grant No. CSD2007-00042, 
by Grant No. PROMETEO/2008/069, 
and by Contract No. MRTN-CT-2006-035482.

\end{document}